\documentclass[conference]{IEEEtran}
\IEEEoverridecommandlockouts
\usepackage{cite}
\usepackage{amsmath,amssymb,amsfonts}
\usepackage{bm}
\usepackage{algorithmic}
\usepackage{graphicx}
\usepackage{textcomp}
\usepackage{xcolor}
\usepackage{tcolorbox}
\usepackage{tikz} %
\usepackage{multirow}
\usepackage{url}
\usepackage{graphicx}
\usepackage{wrapfig}
\usepackage{subcaption}
\usepackage{tcolorbox}
\usepackage{mathtools}
\usepackage{adjustbox}
\usepackage{array}
\usepackage{blindtext}
\usepackage{stfloats}
\usepackage{flushend}

\def\BibTeX{{\rm B\kern-.05em{\sc i\kern-.025em b}\kern-.08em
    T\kern-.1667em\lower.7ex\hbox{E}\kern-.125emX}}
\begin{document}

\title{MinGRU-Based Encoder for Turbo Autoencoder Frameworks\\
{}
\thanks{{This work was supported in part by the German Research Foundation (DFG) as part of Germany’s Excellence Strategy - EXC 2050/1 - Project ID 390696704 - Cluster of Excellence \emph{``Centre for Tactile Internet with Human-in-the-Loop’' (CeTI)} of Technische Universität Dresden and by the German Federal Ministry of Education and Research (BMBF) within the national initiative on 6G Communication Systems through the research hub \emph{6G-life} under Grant 16KISK001K as well as the 6G-ANNA project under grant 16KISK103.}}
}

\author{
    \IEEEauthorblockN{Rick Fritschek and Rafael F. Schaefer\\}
    \IEEEauthorblockA{\centering
    \begin{tabular}{c}
    \textit{Chair of Information Theory and Machine Learning }\\
    \textit{Technische Universit\"at Dresden}\\
      Dresden, Germany\\
    \texttt{\{rick.fritschek, rafael.schaefer\}@tu-dresden.de}
	\end{tabular}
    }}

\maketitle

\begin{abstract}
Early neural channel coding approaches leveraged dense neural networks with one-hot encodings to design adaptive encoder-decoder pairs, improving block error rate (BLER) and automating the design process. However, these methods struggled with scalability as the size of message sets and block lengths increased. TurboAE addressed this challenge by focusing on bit-sequence inputs rather than symbol-level representations, transforming the scalability issue associated with large message sets into a sequence modeling problem. While recurrent neural networks (RNNs) were a natural fit for sequence processing, their reliance on sequential computations made them computationally expensive and inefficient for long sequences. As a result, TurboAE adopted convolutional network blocks, which were faster to train and more scalable, but lacked the sequential modeling advantages of RNNs. 
Recent advances in efficient RNN architectures, such as minGRU and minLSTM, and structured state space models (SSMs) like S4 and S6, overcome these limitations by significantly reducing memory and computational overhead. These models enable scalable sequence processing, making RNNs competitive for long-sequence tasks.
In this work, we revisit RNNs for Turbo autoencoders by integrating the lightweight minGRU model with a Mamba block from SSMs into a parallel Turbo autoencoder framework. Our results demonstrate that this hybrid design matches the performance of convolutional network-based Turbo autoencoder approaches for short sequences while significantly improving scalability and training efficiency for long block lengths. This highlights the potential of efficient RNNs in advancing neural channel coding for long-sequence scenarios.
\end{abstract}


\section{Introduction}
Practical channel codes enable reliable message reconstruction over noisy or nonlinear media like air or optical fiber using low-complexity decoders that achieve low block error rates (BLER) and high code rates \cite{LinandCostello, Gallagerbook}. While capacity-achieving polar codes \cite{Arikan} are optimal for binary-input memoryless symmetric channels, practical design requires adapting encoder-decoder pairs to varying blocklengths, rates, and channel parameters \cite{SCLPolar,UrbankeLDPCNotStable,ArikanPACArxiv}. Changes in non-standard channel models can degrade performance, necessitating automated code design. To address this, (deep) neural networks have been proposed as fully learned and adaptive encoder-decoder alternatives, improving BLER and complexity compared to classic codes. The initial work used dense neural networks in conjunction with one-hot encoded signals and cross-entropy loss \cite{DeepLearningEnd2End}. This was a remarkable success, as it demonstrated the potential to jointly learn all required signal processing and coding tasks, presenting a paradigm shift in communication system design. However, it quickly became apparent that this approach suffered from scaling limitations, as large block lengths $n$ and message sets $k$ would lead to an exponential increase in nodes, as one-hot nodes scale with $2^k$. Furthermore, experiments show that embeddings also have inherent scaling problems. There are many approaches to remedy these scaling issues: \cite{CCNCodes} use a concatenation of classical RS codes with the neural network codes to boost the block lengths by a multiplicative factor while still maintaining the symbol-wise input and cross-entropy loss, which is beneficial to optimize for block error rate; \cite{ProductAE} utilizes smaller encoder networks and combines them in a Cartesian multiplication fashion to get a better scaling model, capable of handling long block lengths; \cite{TurboAE}, which uses the Turbo principle together with neural network blocks and binary input; \cite{TenBrinkSerialTurboAE} a variant of the aforementioned idea which uses a serial structure instead of a parallel structure in the Turbo-autoencoder design of the encoder-decoder blocks, enabling unfolding and component-wise training via Gaussian priors. These works view the input symbols as a sequence of bits, which are then processed sequentially in the encoder, therefore circumventing the scaling problem from before, and shifting the focus to how best to handle long sequences and memory. Then the question arises which neural network architecture to use for the blocks inside the turbo structure. Recurrent neural networks, and in particular the variants LSTM (Long Short Term Memory) and GRU (Gated Recurrent Unit), are natural candidates for sequential processing of messages,  as they are widely used for sequential data in machine learning, e.g.\cite{karpathy2015rnn, sutskever2014sequence} and are universal approximators \cite{schafer2006recurrent}.  GRU decoders have also been shown to achieve optimal decoding performance for classical codes \cite{kim2018communication}. However, due to their nature of processing, they require sequential computation, and therefore gradient backpropagation through time during training. This leads to training times that are linear in sequence length, which limits the practicability for long to very long sequences. The empirical result is that using GRUs for the Turbo Autoencoder needs 10x more GPU memory while being 10x slower to train \cite{TurboAE}. Therefore, subsequent works focused on convolutional neural network based blocks, which are faster to train. However, recently there is renewed interest in efficient RNNs and state space models both of which aim to address the inefficiencies of traditional RNNs. These newer developments aim to combine the sequential processing capabilities of RNNs with improved computational and memory efficiency, making them viable for longer sequences and higher throughput requirements. Key advancements include: 1.) State Space Models (SSMs): Recent models like S4 (Structured State Space for Sequence Modeling) \cite{gu2022efficiently} and its variants propose replacing traditional recurrent operations with structured linear systems that can process sequences in parallel while retaining the ability to model long-range dependencies. By leveraging the mathematical properties of state space representations, these models reduce the need for sequential backpropagation, offering significant speed and memory improvements. These efforts culminated in the recently introduced S6 model \cite{gu2023mamba}, which uses a specific Mamba block around the SSM core, gaining considerable attention due to its strong performance on sequential tasks. 2.) Work on parallelization of RNNs: where \cite{orvieto2023resurrecting} linearized and diagonalized the recurrence; whereas \cite{beck2024xlstm} built on improving the LSTM model; and \cite{feng2024were} introduced a new minimal version of GRUs and LSTMs, called minLSTM and minGRU.

In this work, we analyze the minGRU model for its use in the Turbo autoencoder structure. For that we combine it with the Mamba block for SSMs and integrate it into the parallel Turbo structure of the encoder. We will show that it matches and at times outperforms the convolutional network performance, while being as fast to train as the convolutional networks for smaller sequence length, and having significant advantages at long block lengths.

\paragraph{Notation}
In this work, we distinguish between random vectors in the communication model and deterministic vectors in the neural network: an $n$-dimensional random vector is denoted by an uppercase letter with a superscript, e.g.\ $X^n$, whose components are $X_1, X_2, \dots$, and whose realized value is $x^n$. In contrast, neural-network vectors are written in bold lowercase with a time subscript, e.g.\ $\mathbf{x}_t$, to represent the input at time $t$ in a recurrent model. For example, $X^n$ corresponds to a random codeword of length $n$ transmitted over the channel (Section~\ref{sec:problemDef}), whereas $\mathbf{x}_t$ is the deterministic input signal at time $t$ in the MinGRU-based block (Section~\ref{sec:MinGRUBlock}).

\section{MinGRU-based Block}
\label{sec:MinGRUBlock}

\subsection{Introduction to Recurrent Neural Networks and Gated Recurrent Units}

Recurrent Neural Networks (RNNs) are a class of neural networks designed to process sequential data by maintaining a hidden state that captures information about previous inputs. Unlike feedforward neural networks, RNNs share parameters across time steps, enabling them to model temporal dependencies effectively. Formally, given an input sequence \( \{ \mathbf{x}_t \}_{t=1}^T \), the hidden state \( \mathbf{h}_t \) at time \( t \) is computed as:
\begin{equation}
\mathbf{h}_t = f(\mathbf{h}_{t-1}, \mathbf{x}_t; \theta),
\end{equation}
where \( f \) is a non-linear function (e.g., a combination of matrix multiplications and activation functions), \( \mathbf{h}_{t-1} \) is the hidden state from the previous time step, \( \mathbf{x}_t \) is the input at time \( t \), and \( \theta \) are the network parameters.

One of the key challenges in training RNNs is the vanishing/exploding gradient problem, which limits their ability to capture long-term dependencies. Gated architectures like the Gated Recurrent Unit (GRU) were introduced to address this limitation.

\subsection{Gated Recurrent Unit (GRU)}
The GRU is a simplified yet effective variant of the Long Short-Term Memory (LSTM) network. It uses gating mechanisms to control the flow of information, which helps in preserving long-term dependencies and mitigating the vanishing gradient issue. The GRU maintains a single hidden state \( \mathbf{h}_t \), updated as follows:
\begin{enumerate}
    \item \textbf{Update Gate}: The update gate \( \mathbf{z}_t \) determines how much of the past hidden state \( \mathbf{h}_{t-1} \)   should be retained:
    \begin{equation*}
    \mathbf{z}_t = \sigma(\mathbf{W}_z \mathbf{x}_t + \mathbf{U}_z \mathbf{h}_{t-1} + \mathbf{b}_z),
    \end{equation*}
    where \( \sigma \) is the sigmoid activation function, and \( \mathbf{W}_z, \mathbf{U}_z, \mathbf{b}_z \) are learnable parameters.

    \item \textbf{Reset Gate}: The reset gate \( \mathbf{r}_t \) controls how much of the past hidden state should influence the current computation:
    \begin{equation*}
    \mathbf{r}_t = \sigma(\mathbf{W}_r \mathbf{x}_t + \mathbf{U}_r \mathbf{h}_{t-1} + \mathbf{b}_r).
    \end{equation*}

    \item \textbf{Candidate Hidden State}: A candidate hidden state \( \tilde{\mathbf{h}}_t \) is computed using the reset gate:
    \begin{equation*}
    \tilde{\mathbf{h}}_t = \tanh(\mathbf{W}_h \mathbf{x}_t + \mathbf{U}_h (\mathbf{r}_t \odot \mathbf{h}_{t-1}) + \mathbf{b}_h), 
    \end{equation*}
    where \( \odot \) denotes the element-wise product.

    \item \textbf{Final Hidden State}: The final hidden state \( \mathbf{h}_t \) is a convex combination of the previous hidden state and the candidate hidden state, weighted by the update gate:
    \begin{equation*}
    \mathbf{h}_t = (1 - \mathbf{z}_t) \odot \mathbf{h}_{t-1} + \mathbf{z}_t \odot \tilde{\mathbf{h}}_t.   
    \end{equation*}
\end{enumerate}
Neglecting the bias terms, and streamlining the matrices as linear operations $\mathrm{Linear}_{d_h}$ with a certain matching hidden dimension $d_h$, a GRU cell has the following simplified form: 
        \begin{align*}
            \bm{z}_t &= \sigma(\mathrm{Linear}_{d_h}([\bm{x}_t, \bm{h}_{t-1}])) \\ 
            \bm{r}_t &= \sigma(\mathrm{Linear}_{d_h}([\bm{x}_t, \bm{h}_{t-1}])) \\ 
            \tilde{\bm{h}}_t &= \mathrm{tanh}(\mathrm{Linear}_{d_h}([\bm{x}_t, \bm{r}_t \odot \bm{h}_{t-1}])) \\
            \bm{h}_t &= (\bm{1} - \bm{z}_t) \odot \bm{h}_{t-1} + \bm{z}_t \odot \tilde{\bm{h}}_t.
        \end{align*}

\subsection{Simplifying GRUs: Introducing minGRU}
The GRU architecture is simplified in two steps to improve efficiency and enable parallel training:
\begin{enumerate}
    \item \textbf{Remove Hidden State Dependencies:} 
    The dependencies on the previous hidden state (\( \mathbf{h}_{t-1} \)) are removed from the update gate (\( \mathbf{z}_t \)) and candidate state (\( \tilde{\mathbf{h}}_t \)), eliminating the need for the reset gate (\( \mathbf{r}_t \)). Specifically, the new equations are:
    \begin{equation*}
    \mathbf{z}_t = \sigma(\text{Linear}_{d_h}(\mathbf{x}_t)), \quad \tilde{\mathbf{h}}_t = \tanh(\text{Linear}_{d_h}(\mathbf{x}_t)).
    \end{equation*}
    By removing hidden state dependencies, all inputs can be computed in parallel, enabling efficient parallel scan-based computation.

    \item \textbf{Remove Range Restrictions on Candidate States:}
    The range restriction imposed by the hyperbolic tangent (\( \tanh \)) on the candidate state (\( \tilde{\mathbf{h}}_t \)) is removed, simplifying its computation to:
    \begin{equation*}
    \tilde{\mathbf{h}}_t = \text{Linear}_{d_h}(\mathbf{x}_t).
    \end{equation*}
\end{enumerate}

Combining these two steps results in the minimal GRU (minGRU), defined as:        
        \begin{align*}
            \bm{h}_t &= (\bm{1} - \bm{z}_t) \odot \bm{h}_{t-1} + \bm{z}_t \odot \tilde{\bm{h}}_t \\
            \bm{z}_t &= \sigma(\mathrm{Linear}_{d_h}(\bm{x}_t)) \\ 
            \tilde{\bm{h}}_t &= \mathrm{Linear}_{d_h}(\bm{x}_t) \\
        \end{align*}
The minGRU reduces parameters from \( O(3d_h(d_x + d_h)) \) in standard GRUs to \( O(2d_h d_x) \), achieving significant efficiency improvements. Parallelization through the parallel scan algorithm results in notable training speedups, with a 175× speedup observed for sequence lengths of 512 on a T4 GPU \cite{feng2024were}. Note that these numbers are based on non-cuDNN optimized implementations.

\section{Problem Definition and General Neural Network Structure}
\label{sec:problemDef}
For testing our proposed encoder structure, we consider a standard point-to-point channel model from the literature. In this model, a binary message $u^k \in \{0,1\}^k$, selected uniformly, is encoded into a real-valued codeword $x^n \in \mathbb{R}^n$ under a power constraint $\mathbb{E}[\|X^n\|^2] \leq nP$. The codeword is transmitted over a noisy channel with output $y^n \in \mathbb{R}^n$. The decoder maps $y^n$ to an estimate $\widehat{u}^k \in \{0,1\}^k$. The channel code rate is $R = k/n$, and performance is measured using the bit error rate (BER) and block error rate (BLER):
\begin{align}
    &\text{BER} = \frac{1}{k} \sum_{i=1}^k \Pr[U_i \neq \widehat{U}_i], \quad  
    \text{BLER} = \Pr[U^k \neq \widehat{U}^k].
\end{align}

We consider the AWGN channel, where $Y^n = X^n + Z^n$, with $Z^n$ being i.i.d. Gaussian noise with zero mean and variance $\sigma^2 = N_0/2$. Assuming that $\|X^n\|^2 \leq n$, the signal-to-noise ratio is $\text{SNR} = 1/N_0$, and $E_b/N_0 = 1/(N_0R)$, where $E_b$ is the energy per information bit.

\subsection{Neural Channel Encoder}

\begin{figure}
    \centering
    \includegraphics[width=0.8\linewidth]{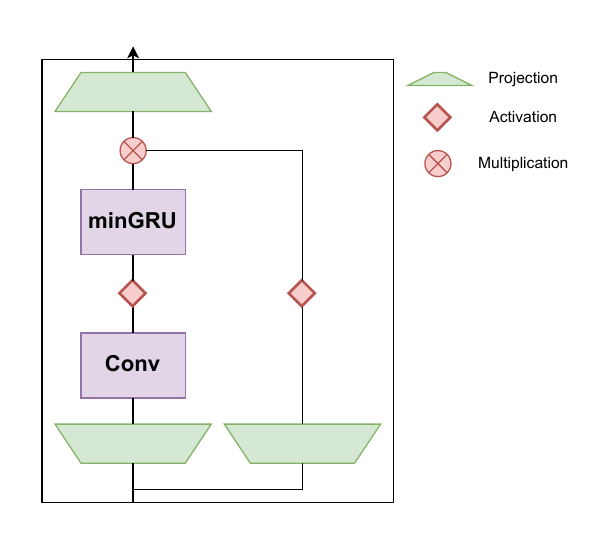}
    \caption{Mamba block with minGRU module. The activations are chosen to be Swish/SILU activations. The projection/transformation upscale the input to the hidden size, which is a parameter and is mostly chosen to be $<12$ here. The multiplication is a multiplicative gate which can be thought as an adaptively learned element-wise filter operation.}
    \label{fig:mamba_gru_block}
\end{figure}

The neural encoder maps the binary input $\tilde{\mathbf{u}} \in \{0,1\}^{k}$ to a real-valued codeword $\tilde{\mathbf{x}} \in \mathbb{R}^{n}$ using a parametric function $f_{\bm{\theta}}$. This forms an $(n, k)$ neural code, where $n \geq k > 0$. For its general structure, we adopt the interleaved encoding structure from \cite{TurboAE}, which processes two strands of the input signal. One strand is directly processed by an encoder block, while the other strand is interleaved and then processed by another encoder block. These blocks therefore operate in parallel, in contrast to the serial processing structure in \cite{TenBrinkSerialTurboAE}. Both outputs of the encoder blocks are then concatenated, and processed by a power constraint. The power constraint is enforced via normalization:
\begin{equation}
\tilde{\mathbf{x}} = \frac{\mathbf{x} - \bm{\mu}_{\mathbf{x}}}{\sigma_{\mathbf{x}}},
\end{equation}
where $\bm{\mu}_{\mathbf{x}}$ and $\sigma_{\mathbf{x}}$ are the mean and standard deviation of $\mathbf{x}$. The normalized codeword $\tilde{\mathbf{x}}$ is transmitted through the AWGN channel.

For each encoder block, we utilize the Mamba module/block Fig.~\ref{fig:mamba_gru_block}, which has a transformation or embedding onto a higher hidden dimension \texttt{F}. To be specific, for an input of \texttt{(B, T, 1)}, where \texttt{B} is the batch size and \texttt{T} the sequence length, the transformation maps the input to \texttt{(B, T, F)}. Subsequently, the sequence is processed by a convolutional layer\footnote{It might seem counter-intuitive at first, to replace a convolutional neural network (CNN) block with another block that includes a CNN, but the CNNs are not comparable in parameters or functionality.} which has the same output dimensionality \texttt{F} as the input and a kernel size of $3$, capturing local patterns over three adjacent sequence steps. Afterward, the signal goes through an SiLU activation which is defined as $\mathrm{SiLU}:=x\sigma(x)$, where $\sigma(x)$ is the sigmoid function, which offers advantages in numerical stability and expressive power and is used for example in \cite{gu2023mamba, beck2024xlstm}. The signal is then processed by the minGRU layer, as explained in the last section, and then merged with a second strand that introduces a gating / filtering mechanism. The gating signal in the second strand also goes through a $\mathrm{SiLU}$ activation, before being multiplicative combined with the minGRU output.  At the end, the signal sequence is transformed down again to input size, i.e. \texttt{(B, T, 1)}. A residual connection around the whole block is included for improved learning stability, allowing the gradient to integrate the block gradually during training.
For our encoder block, we stack two of these Mamba-minGRU sub-blocks. At the end, as in the general Turbo autoencoder structure, two of these encoder blocks (which consist of two stacked Mamba-minGRU sub-blocks each) are concatenated in the last dimension yielding the encoder output signal \texttt{(B, T, 2)} which is normalized and sent over the channel.

\subsection{Neural Decoder}

The neural decoder, denoted by $g_{\bm{\theta}}$, maps the received sequence $\mathbf{y} \in \mathbb{R}^n$ to an estimated binary message $\widehat{\mathbf{u}} \in \{0,1\}^k$. In the final layer of the network, we apply a sigmoid function $\sigma_{\text{sigmoid}}(z) = 1/(1 + e^{-z})\in [0,1] $
and we train the decoder using the standard binary cross-entropy (BCE) loss:
\[
-\frac{1}{k} \sum_{i=1}^{k} \Bigl[
    u_i \,\log(\hat{p}_i)
    \;+\;
    \bigl(1 - u_i\bigr)\,\log\bigl(1 - \hat{p}_i\bigr)
\Bigr],
\]
where $u_i \in \{0,1\}$ is the ground-truth bit and $\hat{p}_i \in [0,1]$ is the predicted probability for bit $i$.

We adopt the parallel Turbo autoencoder (Turbo-AE) decoder structure, following \cite{jiang2019deepturbo} and \cite{TurboAE}, which has demonstrated strong standalone performance. For a rate $0.5$ setup, the input signal is split into two parts: $\mathbf{a}$ (systematic) and $\mathbf{b}$ (parity),
and interleaving is applied to form $\mathbf{a}_{\text{int}}$ and $\mathbf{b}_{\text{int}}$, while a corresponding deinterleaver yields $\mathbf{b}_{\text{deint}}$.

The decoder proceeds iteratively in two stages (two CNN blocks) per decoding iteration:
\begin{enumerate}
    \item \emph{First CNN Block.}
    It processes the concatenation $[\,\mathbf{a},\;\mathbf{b}_{\text{deint}},\;\mathbf{prior}\,]$
    where $\mathbf{prior}$ is the prior log-likelihood ratio (LLR) estimate from the previous iteration (initialized to zero in the first iteration). The CNN block outputs 
    $\mathbf{x}_{\text{dec}}$ and $\mathbf{x}_{\text{plr}}$,
    where $\mathbf{x}_{\text{dec}}$ is an intermediate decoded feature map and $\mathbf{x}_{\text{plr}}$ are predicted LLRs. The \emph{extrinsic information} is then computed as
    \[
      \mathbf{x}_{\text{ext}} \;=\; \mathbf{x}_{\text{plr}} \;-\; \mathbf{prior}.
    \]
    This extrinsic information is interleaved to form $\mathbf{x}_{\text{ext},\text{int}}$.

    \item \emph{Second CNN Block.}
    The second CNN block processes $[\,\mathbf{a}_{\text{int}},\;\mathbf{b},\;\mathbf{x}_{\text{ext},\text{int}}\,]$.
It produces updated LLRs $\mathbf{x}_{\text{plr}}$, from which we form a new $\mathbf{prior}$ by deinterleaving the residual:
    \[
      \mathbf{prior}
      \;=\;
      \mathrm{deinterleave}\bigl[\mathbf{x}_{\text{plr}} \;-\; \mathbf{x}_{\text{ext},\text{int}}\bigr].
    \]
    This completes a single iteration of the Turbo decoding procedure.
\end{enumerate}

Each CNN block contains $L$ convolutional layers, each with kernel size $k$ and $h$ channels. In our experiments, we choose $k=5$ and $h=100$, and vary the number of layers $L$ between $2$ and $5$. Compared to the standard Turbo-AE, we employ non-linear activations (e.g., ELU) in each convolutional layer, which gave slight performance improvements. We use standard zero-padding to preserve sequence length; circular padding did not show measurable gains and required more computation time.
We perform up to $6$ iterations of this two-block decoding procedure. After the final iteration, the LLR outputs are passed through the sigmoid activation to obtain the final bit estimates $\widehat{\mathbf{u}}$.

\section{Simulations and Results}


\begin{figure}
    \centering
    \includegraphics[width=0.9\linewidth]{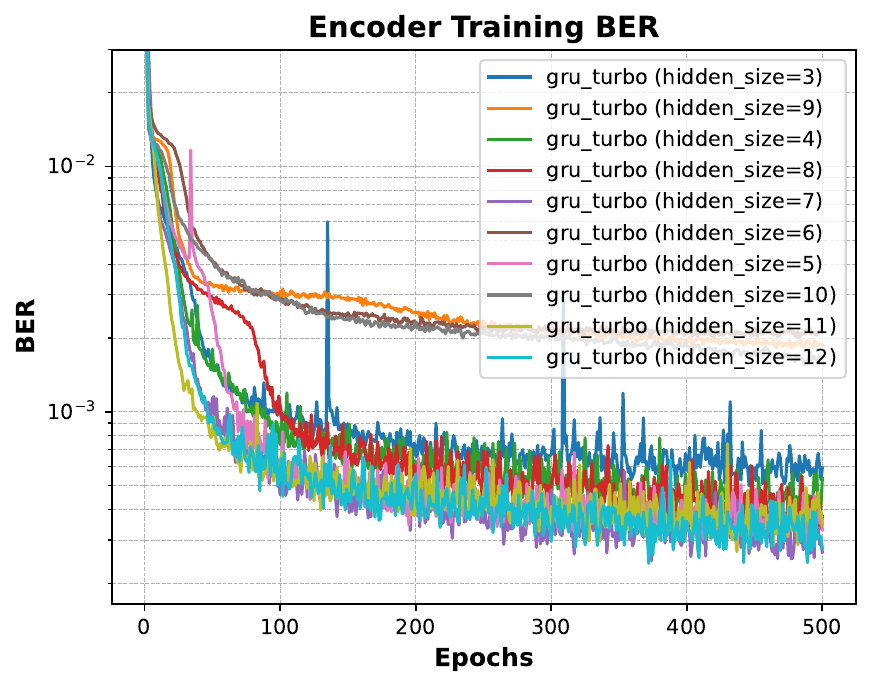}
    \caption{Experiments for various feature sizes for the minGRU encoder blocks. Showcasing the behavior of the BER for encoder training over $500$ epochs.}
    \label{fig:hidden_size_ber}
\end{figure}

\begin{figure}
    \centering
    \includegraphics[width=0.9\linewidth]{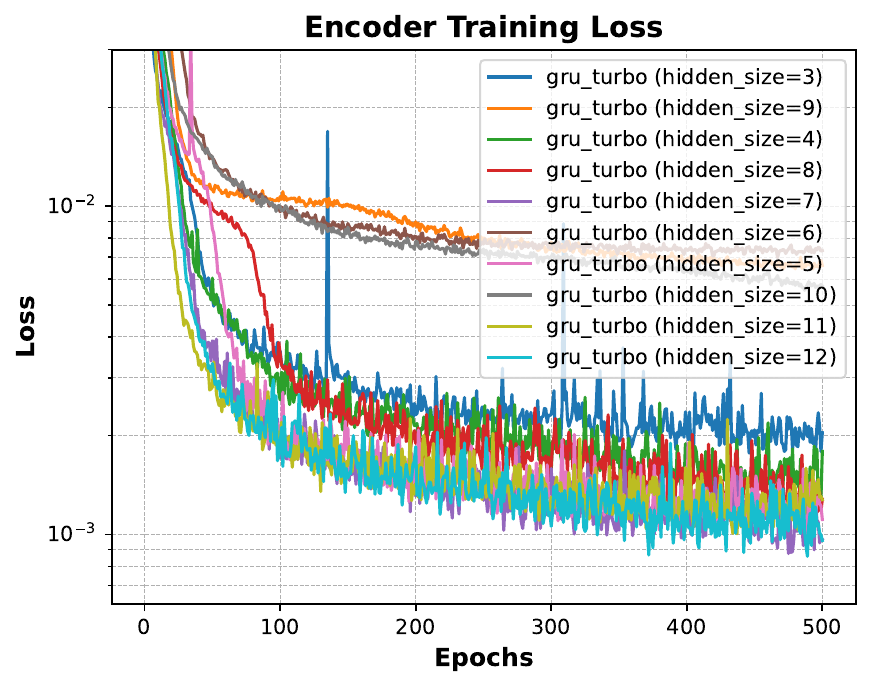}
    \caption{Experiments for various feature sizes for the minGRU encoder blocks. Showcasing the behavior of the training loss for encoder training over $500$ epochs.}
    \label{fig:hidden_size_bler}
\end{figure}

For the simulations, our general setup\footnote{Our code will be made available under https://github.com/Fritschek} was that we tested all frameworks / codes with a message length of $64$ and a block length of $128$ for $500$ epochs, and a sample size of $50000$ per epoch. The product encoder was chosen to be of size $(8 \cdot 16,8\cdot 8)$, which we trained for a batch size of $5000$, and for $5000$ epochs, to get close to the baseline from \cite{ProductAE}.
For our setup, we use a batch size of $128$ for the encoder, and a batch size of $512$ for the decoder. Moreover, the decoder is trained five times more frequently than the encoder. This asymmetric batch size of our encoder-decoder setup, as well as for the CNN turbo autoencoder setup and the serial CNN turbo autoencoder setup, empirically showed improved performance for all codes that we tested. We also refrained from too much fine-tuning, i.e., we kept the batch size constant over time, and the learning rate at $2\times10^{-4}$ for better comparability. For optimization we used the AdamW optimizer \cite{loshchilov2018decoupled} with a standard weight decay of $0.01$.   Our model and the two turbo autoencoder frameworks had $5$ decoder layers and $2$ encoder layers, while the product autoencoder had $7$ encoder layers and $9$ decoder layers. For the feature size, or information flow dimension within the frameworks, we followed closely the guidelines from the papers, i.e., $F=5$ for the parallel turbo autoencoder, $F=10$ for the serial autoencoder and $F=3$ for the product autoencoder. Our model, using the parallel turbo decoder model, also utilized a feature size of $5$ in the decoder. For our hidden size or feature size in the encoder, we used $F=4$. Note that we tested various feature sizes for our minGRU model, see Fig.~\ref{fig:hidden_size_ber} and Fig.~\ref{fig:hidden_size_bler}. The results were mostly consistent, converging to roughly the same level. Computational time was also similar, increasing to roughly double for drastically higher feature dimension such as $F=128$, while maintaining the same result range. We therefore opted for $F=4$ for the following experiments.

\begin{figure*}[t!]
    \centering
    \begin{subfigure}[b]{0.32\textwidth}
        \centering
        \includegraphics[width=1\linewidth]{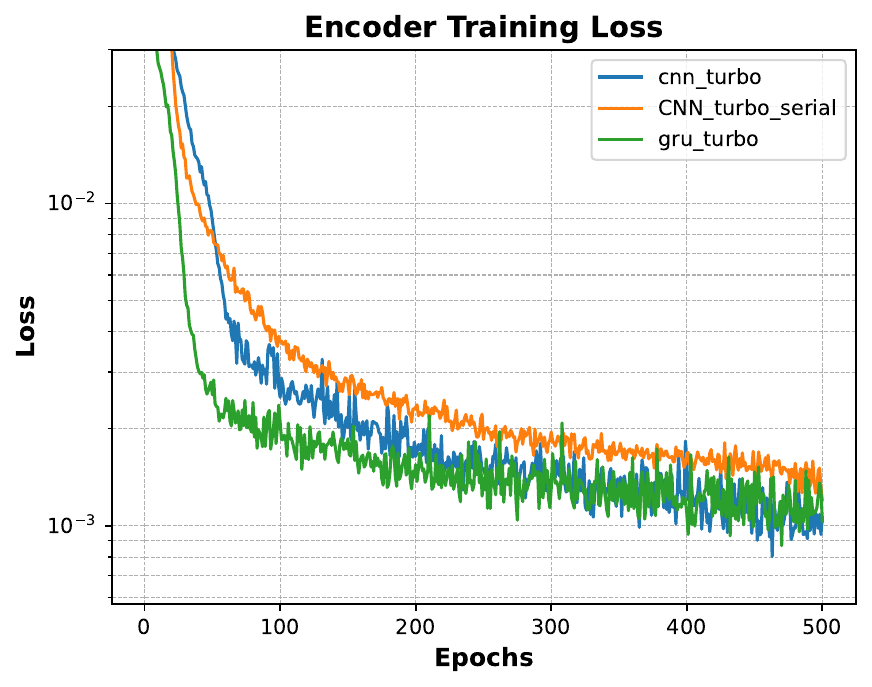}
        \caption{Training loss development for our MinGru model, as well as the parallel and the serial Turbo autoencoder.}
        \label{fig:Training Loss}
    \end{subfigure}
    \begin{subfigure}[b]{0.32\textwidth}
        \centering
        \includegraphics[width=\linewidth]{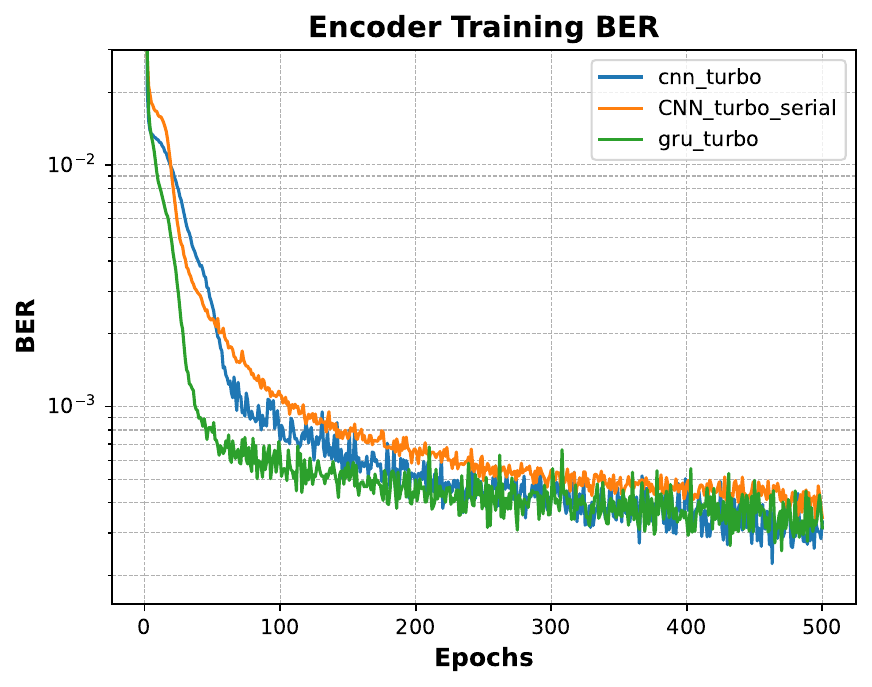}
        \caption{Training BER development for our MinGRU model, and the two Turbo autoencoder architectures.}
        \label{fig:Training BER}
    \end{subfigure}
    \begin{subfigure}[b]{0.32\textwidth}
        \centering
        \includegraphics[width=\linewidth]{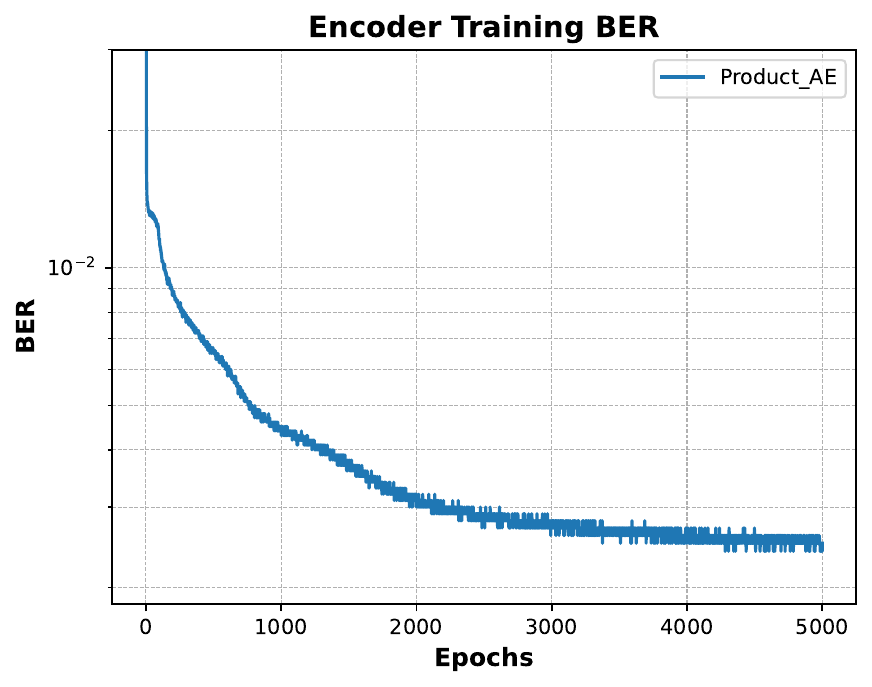}
        \caption{Training BER development for the Product autoencoder over $5000$ epochs, and therefore separately plotted.}
        \label{fig:Training Ber Product}
    \end{subfigure}
    \caption{Training dynamics for the tested models.}
    \label{fig:three-figures}
\end{figure*}

\begin{figure}
    \centering
    \includegraphics[width=0.9\linewidth]{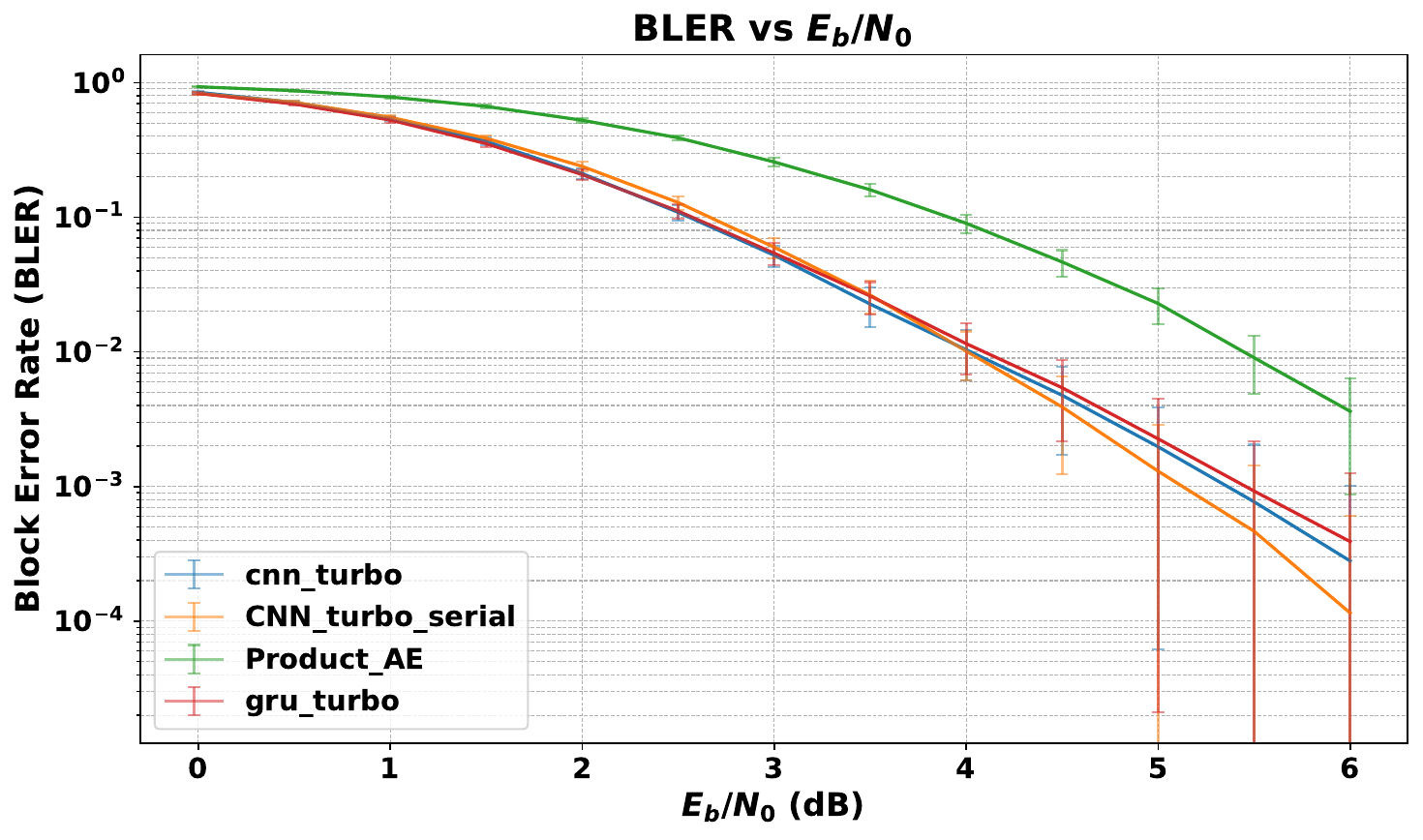}
    \caption{Bler values for various models. All models were tested for $N=50000$ samples per $E_b/N_0$ point until $E_b/N_0=3.5$~dB, and $2N$ samples from $4$~dB, $3N$ from $5$~dB and $4N$ from $5.5$~dB.}
    \label{fig:Bler Results}
\end{figure}

\begin{figure}
    \centering
    \includegraphics[width=0.9\linewidth]{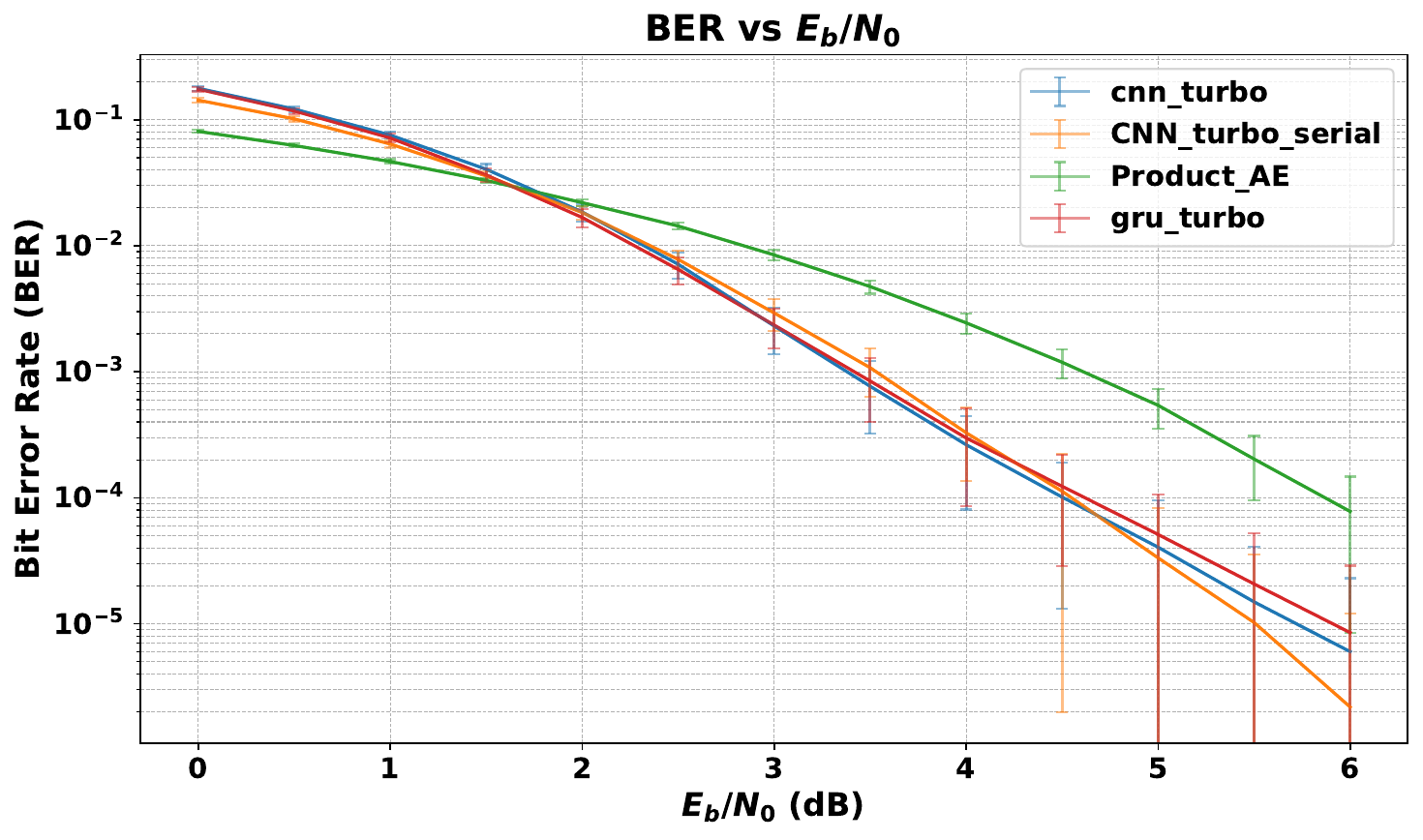}
    \caption{Ber values for various models. All models were tested for $N=50000$ samples per $E_b/N_0$ point until $E_b/N_0=3.5$~dB, and $2N$ samples from $4$~dB, $3N$ from $5$~dB and $4N$ from $5.5$~dB.}
    \label{fig:BER Results}
\end{figure}

For all models, we employed alternate encoder-decoder training, where the decoder is trained five times more frequently than the encoder, as is standard in the literature. Combined with the previously described asymmetric batch sizes ($128$ for the encoder and $512$ for the decoder) and a total of $50000$ samples per epoch, the encoder was trained on approximately $390$ batches per epoch (each containing $128$ samples from those $50000$), while the decoder was trained on about $1950$ batches per epoch (each with $512$ samples). Consequently, the decoder processed 20 times more samples than the encoder.
Fig.~\ref{fig:Training BER} and Fig.~\ref{fig:Training Loss} show that the MinGRU exhibits slightly better training behavior than state-of-the-art models, converging more quickly to a stable plateau. 
Notably, our simulations of the serial turbo autoencoder and the product autoencoder achieved slightly lower performance compared to the best results reported in \cite{TenBrinkSerialTurboAE} and \cite{ProductAE}. This performance gap could stem from the absence of fine-tuning or nuances in their approach that were not explicitly described, as the original implementations are not publicly available.
In Fig.~\ref{fig:Bler Results} and Fig.~\ref{fig:BER Results}, we present the inference results, showing the achieved block error rate (BLER) and bit error rate (BER), respectively. To reduce variance at higher \(E_b/N_0\) levels, we simulated with an increasing sample size. However, the highest \(E_b/N_0\) points still displayed significant variance and error, prompting us to include error bars for full transparency.
The results show that the MinGRU and the two turbo autoencoder models perform similarly, with their results falling within error bounds. In contrast, the product autoencoder performs slightly worse in terms of BLER and at very high \(E_b/N_0\) for BER but demonstrates slightly better BER performance in the very low \(E_b/N_0\) regime. Additionally, the serial turbo autoencoder appears to generalize better, suggesting that applying the MinGRU architecture to the serial autoencoder could be a worthwhile investigation to assess whether this behavior persists.

\section{Conclusion and Outlook}

We have demonstrated that our proposed novel encoder, which integrates the recently developed minimal GRU algorithm with a Mamba block, achieves competitive performance compared to CNN-based Turbo autoencoder models. Notably, the encoder block outperforms the standard RNN-based versions of Turbo autoencoders \cite{TurboAE} in terms of training loss behavior, showcasing faster convergence and reaching lower error values more quickly. Furthermore, training this architecture is significantly faster than RNN-based counterparts, making it a practical alternative for longer block lengths, as the minimal GRU no longer exhibits linear sequence time dependence. However, while our own current implementation is approximately as fast to train as the CNN variant for short to medium sequence lengths (up to $1000$), we believe that leveraging an optimized cuDNN-based implementation could substantially accelerate the training process. This could make the proposed architecture a more viable option in scenarios where efficiency and scalability are critical. Interestingly, the inclusion of the Mamba structure around the minimal GRU algorithm has proven to be a key innovation, yielding superior results in terms of both training dynamics and overall performance. Given that this approach is still in its early stages, we believe that further fine-tuning and systematic exploration of hyperparameters could unlock even greater potential.

Future work will focus on adapting this setup for the decoder and exploring modifications to optimize its performance. Moreover, it would be valuable to investigate the behavior of a pure minimal GRU setup for very long sequence lengths, as this could reveal new insights into its scalability and robustness.

\bibliographystyle{IEEEtran}
\bibliography{refs}

\end{document}